\documentclass[twocolumn,superscriptaddress,aps,pra]{revtex4-1}

\usepackage{amsmath,amssymb,graphicx,xcolor,bm,soul}
\usepackage{upgreek}
\usepackage{xfrac}
\usepackage{braket}

\usepackage[english]{babel}
\usepackage{natbib}
\definecolor{orange}{RGB}{235, 129, 0}
\usepackage[bookmarks=true,
            bookmarksnumbered=false,
            bookmarksopen=true,
            breaklinks=true,
            pdfborder={0 0 0},
            final=true,
            backref=none,
            colorlinks=true,
            linkcolor=blue,
            menucolor=blue,
            citecolor=blue,
            urlcolor=blue]{hyperref}

\begin{document}

\title{Unsupervised quantum machine learning for fraud detection}

\author{Oleksandr Kyriienko}
\affiliation{Department of Physics and Astronomy, University of Exeter, Stocker Road, Exeter EX4 4QL, United Kingdom}

\author{Einar B. Magnusson}
\affiliation{Markets \& Securities Services, HSBC Bank Plc., 8 Canada Square, London E14 5HQ, United Kingdom}

\begin{abstract}
We develop quantum protocols for anomaly detection and apply them to the task of credit card fraud detection (FD). First, we establish classical benchmarks based on supervised and unsupervised machine learning methods, where average precision is chosen as a robust metric for detecting anomalous data. We focus on kernel-based approaches for ease of direct comparison, basing our unsupervised modelling on one-class support vector machines (OC-SVM). Next, we employ quantum kernels of different type for performing anomaly detection, and observe that quantum FD can challenge equivalent classical protocols at increasing number of features (equal to the number of qubits for data embedding). Performing simulations with registers up to 20 qubits, we find that quantum kernels with re-uploading demonstrate better average precision, with the advantage increasing with system size. Specifically, at 20 qubits we reach the quantum-classical separation of average precision being equal to 15\%. We discuss the prospects of fraud detection with near- and mid-term quantum hardware, and describe possible future improvements.
\end{abstract}

\maketitle

\section{Introduction} Quantum computing (QC) can potentially offer advantage in solving large-scale computational problems \cite{NielsenChuang}. Most likely beneficiaries of QC include chemistry \cite{McArdle2020}, material science \cite{MatSci,Cade2020}, cryptography and secure computing \cite{Wallden2019,Leichtle2021}, high energy physics \cite{Preskill2018}, multisimulation \cite{kyriienko2021solving,lubasch2020variational}, as well as data \cite{Lloyd2014,Biamonte2017,Benedetti2019} and financial \cite{Orus2019a,Herman2022} analysis. Specifically, financial applications of quantum computing include optimization \cite{Orus2019a}, speeding-up Monte-Carlo for derivative pricing \cite{Herman2022,Stamatopoulos2022towardsquantum}, and numerous tasks targeted by quantum machine learning \cite{Benedetti2019,Bharti2022}.

Quantum machine learning (QML) has attracted attention due to the possibility of speeding up various data-driven tasks. Quantum linear algebra subroutines can offer ways of advancing QML in the long run \cite{Biamonte2017}, but are too costly for near-term operation. The rise of variational quantum protocols has brought another wave of ideas in QML \cite{cerezo2021variational}, with the goal shifted towards constructing expressive models as parametrized quantum circuits \cite{Benedetti2019}. Here, the major ingredient corresponds to feature map-based encoding of data, which favors near-term operation \cite{Mitarai2018,Schuld2019QML,Abbas2021}. To date, the most widespread tasks for QML correspond to \emph{supervised} learning, and in particular classification \cite{Schetakis2022,Li2021}. Examples include variational quantum classifiers \cite{Havlicek2019,Mitarai2018,Schuld2020,PerezSalinas2020datareuploading,Nghiem2021}, quantum convolutional neural networks \cite{Cong2019,Zhao2021analyzingbarren,Chen2022}, and quantum kernel-based classification \cite{Havlicek2019,Schuld2019QML,Schuld2021kernel}. Other supervised QML tasks correspond to regression \cite{Mitarai2018,Paine2022}, solving linear \cite{XU20212181,Chen2019a} and differential equations \cite{kyriienko2021solving,lubasch2020variational,Paine2022}, and quantum model discovery \cite{Heim2021}.

Going beyond supervised methods, generative modelling and specifically generative adversarial networks (GANs) attracted a significant attention \cite{PerdomoOrtiz2018,Zeng2019,Zoufal2019,Romero2021,Paine2021}. However, other subjects of unsupervised learning, specifically corresponding to clustering, remain under-explored. Among the few advances are quantum principal component analysis (PCA) \cite{Lloyd2014}, quantum kernel PCA \cite{Liu2018} and q-means (quantum version of k-means clustering) \cite{Kerenidis2019}. These however target fault-tolerant operation in the long term and primarily assume a purely quantum dataset. From the near-term perspective, the prediction of financial crashes was analyzed using quantum annealer architecture \cite{Orus2019} and recently a variational quantum clustering was proposed for the digital operation mode \cite{Bermejo2022}.
%
%%%
\begin{figure}
\centering
\includegraphics[width=0.9\linewidth]{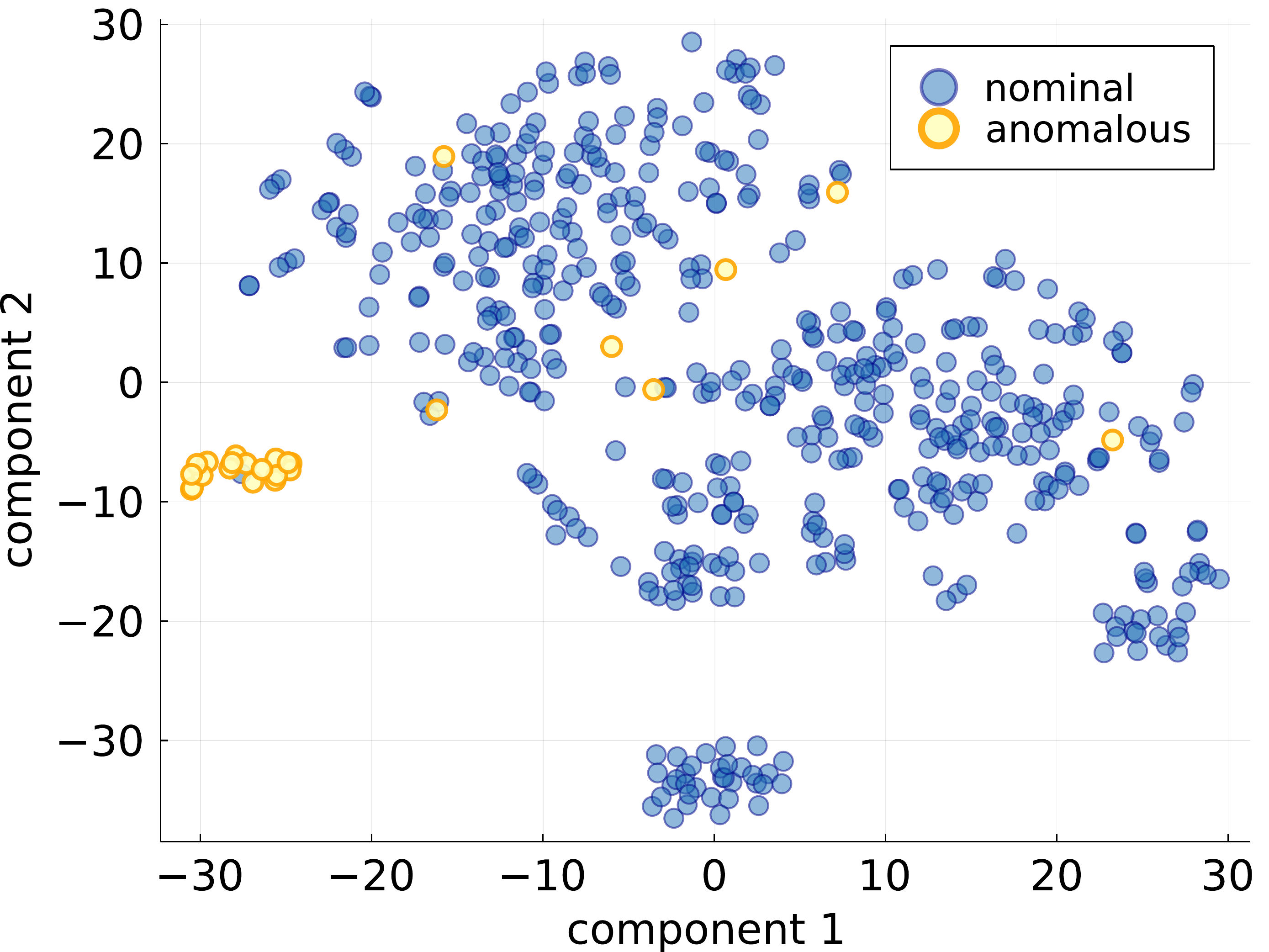}
\caption{\textbf{Data visualization}. We consider a credit card fraud detection dataset labelled by $N=28$ features, and sub-sampled to include 25 fraud cases for $N_s=525$ total. We use t-SNE to present the data with nominal (blue) and anomalous (yellow) classes.}
\label{fig:tSNE}
\end{figure}
%%%

In this work, we target an important QML-for-finance application corresponding to the task of fraud detection (FD). This can be seen as detection of anomalies (outliers) in datasets with dominant number of nominal (genuine) samples \cite{Pang2022}. One example relevant from the business perspective corresponds to credit card fraud detection \cite{BinSulaiman2022}. Essentially, the task corresponds to clustering nominal data in an unsupervised fashion, such that potential fraud events can be labeled as anomalous. The problem becomes challenging when datasets are described by samples with many features, which additionally are strongly correlated. Classical machine learning methods approach anomaly detection in several ways. Distance-based anomaly detection is performed by measuring distances between samples and assigning corresponding clusters \cite{Ramaswamy2000}. Density-based approaches such as local outlier factor (LOF) algorithm \cite{Breunig2000} use estimates of local density, and predict outliers from low-density regions. Isolation Forest \cite{IsolationForest} was proposed as a way of isolating anomalies on binary trees. However, the aforementioned methods do not have direct quantum analogs. Deep anomaly detection was recently approached by utilizing deep neural networks, with a notable example of AnoGAN \cite{Schlegl2017,SCHLEGL201930}, and its quantum version was considered recently in Ref.~\cite{Herr2021}. Finally, kernel-based methods are regularly used for anomaly detection as performed by one-class support vector machines (OC-SVM) \cite{Scholkopf1999} and support vector data description \cite{Tax2004}, which use high-dimensional support vectors for drawing decision boundaries between nominal and anomalous samples. Recently, kernel-based methods were used as an integral part of state-of-the-art anomaly detection \cite{Ruff2018,ZHANG20211}.

Here, we develop a connection between quantum kernel methods and unsupervised anomaly detection. Specifically, we use quantum kernels as a basis for one-class SVM and apply the approach to detect fraud in a credit card transaction dataset. Using pre-processing of the dataset, we test the approach at increasing number of qubits (features), and compare to classical results obtained from optimized radial basis function kernels. We observe that at 20 qubits the separation between quantum and classical detection precision reaches 15\%, highlighting high expressivity of quantum kernels and potential advantage in boosting FD performance. We also discuss potential prospects and open challenges for the area.

\section{Fraud detection: classical benchmarks} We aim to develop strategies for detecting anomalies, specifically considering financial applications in fraud detection. For this, we use a credit card fraud dataset from Kaggle \cite{Kaggle}, where anonymized data is collected into nominal (genuine) and anomalous (fraudulent) transactions as described by 28 anonymized features. These are nameless variables represented by float numbers.  Additionally, two features being ``time'' and ``amount'' of transaction are supplied. For simplicity, we only consider the 28 anonymized features. The dataset contains 492 fraud cases from the total of 284,807 transactions, and is highly imbalanced. In this situation, the accuracy metric (ratio of the number of correct predictions to total number of predictions) should be avoided as it is not sufficiently informative. The F1 score (harmonic mean of recall and precision with respect to the anomaly class) is often used in imbalanced classification problems. However, as this requires setting a classification score threshold, we prefer using a threshold-free metric called average precision (AP). This corresponds to the precision at each threshold, weighted by the increase in recall from previous threshold and averaged over the full grid.
%
%\begin{equation}
%\label{eq:AP}
%\mathrm{AP} = \sum_i (R_i - R_{i-1}) P_i,    
%\end{equation}
%
%where $R_i$ and $P_i$ are values of recall and precision at the threshold $i$.

To proceed with the analysis, we sub-sample the dataset to make it manageable for near-term quantum simulations. We furthermore sample selectively to increase the share of anomalies and balance the dataset. We select 500 samples including 25 fraud examples and visualize the high dimensional dataset via t-SNE. The result is shown in Fig.~\ref{fig:tSNE}. We observe that visually t-SNE shows a separation between nominal and anomalous samples, but sole visualization map does not allow to track all fraudulent transactions, and additional clusters of nominal transactions are also present. This demonstrates the challenge of high-quality fraud detection. 

Next, we perform dimensionality reduction to match the number of features with the number of qubits used in simulation. For this, we use principal component analysis and keep only the first $N$ principal components. Before applying PCA, we use feature-wise standard scaling, i.e. we subtract the mean and scale by the standard deviation for each feature. We also vary the number of features $N$ to see how the classical detection changes with the number of features. Later, we use quantum data loading techniques where $N$ corresponds to the number of qubits, thus facilitating the comparison between classical and quantum methods at increasing system size. We note that in the quantum case it is also important to scale features such that each component can be loaded by using quantum circuits.
%%%
\begin{figure}
\centering
\includegraphics[width=1.\linewidth]{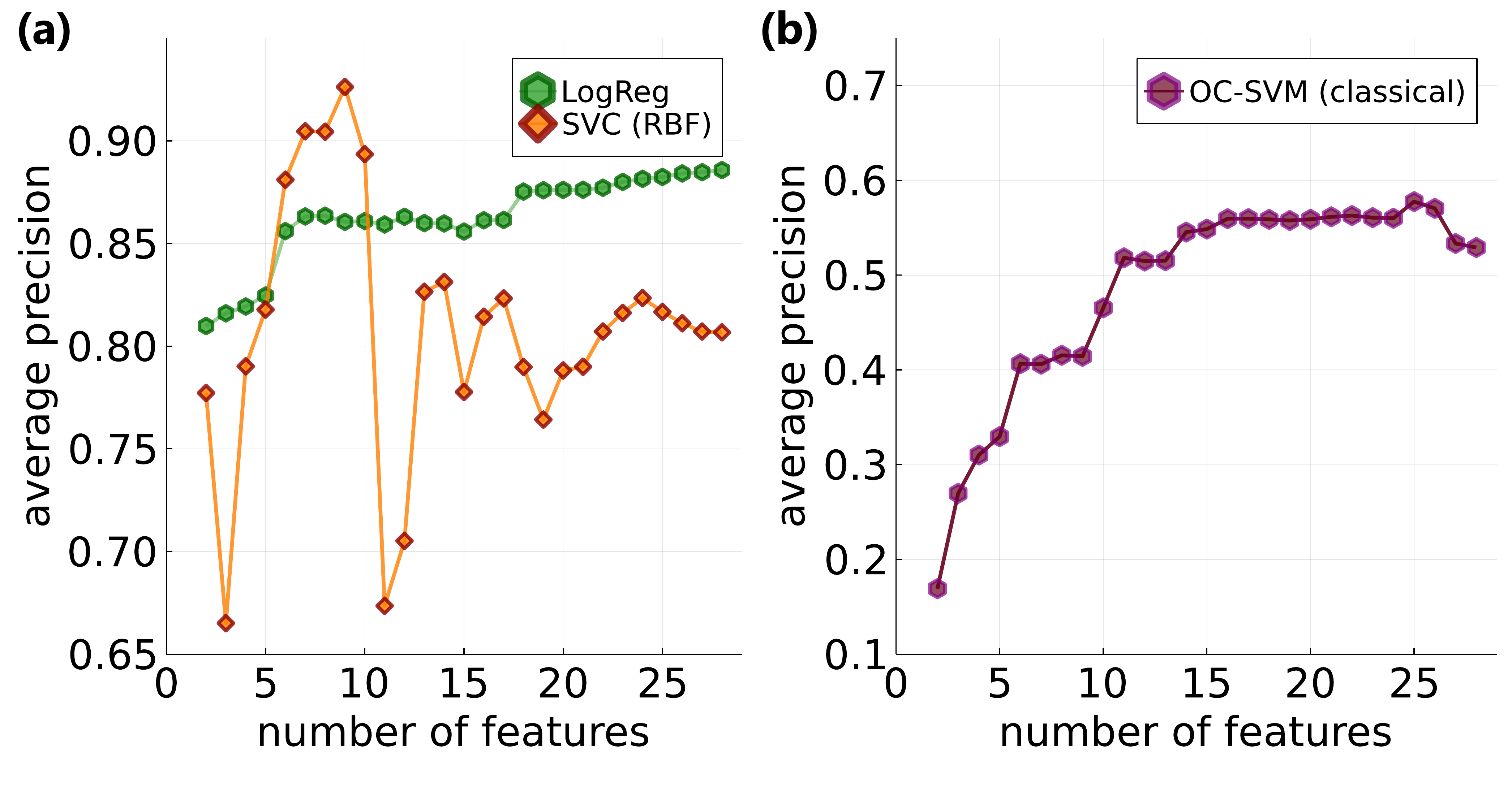}
\caption{\textbf{Classical benchmarking. (a)}. Average precision of fraud detection shown at increased number of used features for supervised fraud detection. We performed supervised ML using logistic regression (LogReg label) and support vector classification (SVC) with RBF kernels. \textbf{(b)} Unsupervised fraud detection with one-class SVM with classical RBF kernels. The model is built considering nominal data only.}
\label{fig:classical}
\end{figure}
%%%

For the classical fraud detection we choose several protocols, which are separated into supervised and unsupervised approaches. The supervised methods include logistic regression and support vector classification (SVC), although the former is simply for added context rather than comparison to a quantum analogue. For the unsupervised method we use one-class support vector machine (OC-SVM). For both SVC and OC-SVM we use radial basis functions (RBF) as the kernel mapping. 

For hyperparameter tuning we choose to tune the regularisation parameter $C$ for all models as well as $\gamma$ in the case of RBF SVM. Hyperparameter tuning is performed by random search and choosing the parameter values that yield the highest cross-validation AP score. Conceptually, this can be troublesome in the unsupervised setting, as target labels are not supposed to be available. However, it is common practice in research on unsupervised methods to consider labels to be present in the validation set, and we adopt the same approach to establish the optimized classical benchmarks.
In the numerical implementations, we use the \textsf{scikit-learn} library for the logistic regression, building SVC models, and performing the hyperparameter search. 

\textbf{Supervised detection (classification).} First, we apply logistic regression (LogReg) with L2 regularization. The results are shown in Fig.~\ref{fig:classical}(a) with green hexagons.  With tuning of $C$ by cross-validation, the performance improves gradually. At the same time, we note that without tuning the performance peaks at five to ten features, and at increased number of features we observed a deterioration of average precision. This shows that models are prone to overfitting at large number of features.

Second, we use a supervised method based on support vector classification \cite{kung2014kernel}. For this, the pre-processed dataset of credit card transactions $\{ \mathbf{x}_i, y_i \}_{i=1}^{N_s}$ is used for building support vectors and drawing a decision boundary via minimizing the hinge loss \cite{Li2021}. The key quantity for building SVC models is a kernel function $\kappa(\mathbf{x}_i, \mathbf{x}_j)$ that quantifies a distance between samples $\mathbf{x}_i$ and $\mathbf{x}_j$ as their inner product. Typically, kernels are designed such that data points are compared in a high-dimensional space. One of the most successful kernels corresponds to a radial basis function (RBF), which is defined as $\kappa(\mathbf{x}_i, \mathbf{x}_j) = \exp (-\gamma \|\mathbf{x}_i - \mathbf{x}_j\|^2)$, where $\gamma$ is a hyperparameter that controls kernel bandwidth, and $\| \cdot \|^2$ denotes squared Euclidean distance between points. The optimal prediction model can be written as $f_{\mathrm{opt}}(\mathbf{x}) = \sum_{i=1}^{N_s} \alpha_i \kappa(\mathbf{x}_i, \mathbf{x})$, where $\alpha_i$ are coefficients coming from the hinge loss minimization. As the model is linear, the optimization is convex, and only requires performing matrix-vector multiplications for the vector of labels $\mathbf{Y}=\{\mathbf{y}_i\}$ and a Gram matrix for data points. The Gram matrix is a matrix $\mathbb{K}(\{ \mathbf{x}_i, \mathbf{x}_j\}_{i,j=1}^{N_s})$ formed by kernels evaluated for all possible pairs of data points. The evaluation of the Gram matrix $\mathbb{K}$ is the most computationally expensive part of SVC, as the number of evaluations scales as $O(N_s^2)$ with the dataset sample size $N_s$, while other post-processing steps are straightforward (implemented in \textsf{scikit-learn}).

The results of the supervised support vector classification are shown in Fig.~\ref{fig:classical}(a) by red diamonds. We observe a level of performance slightly higher than the logistic regression for five-to-ten feature region, but note that the performance at larger number of features drops slightly in AP, likely due to overfitting. These results set the bar from classical benchmarking of supervised classification, which we later compare to quantum kernel approach.
We can conclude that supervised methods overall show high performance, but require prior knowledge of fraud cases and do not generalize to new type of fraud. We proceed to show how to detect previously unseen fraud types with unsupervised methods. 
%%%
\begin{figure}
%\centering
\includegraphics[width=1.\linewidth]{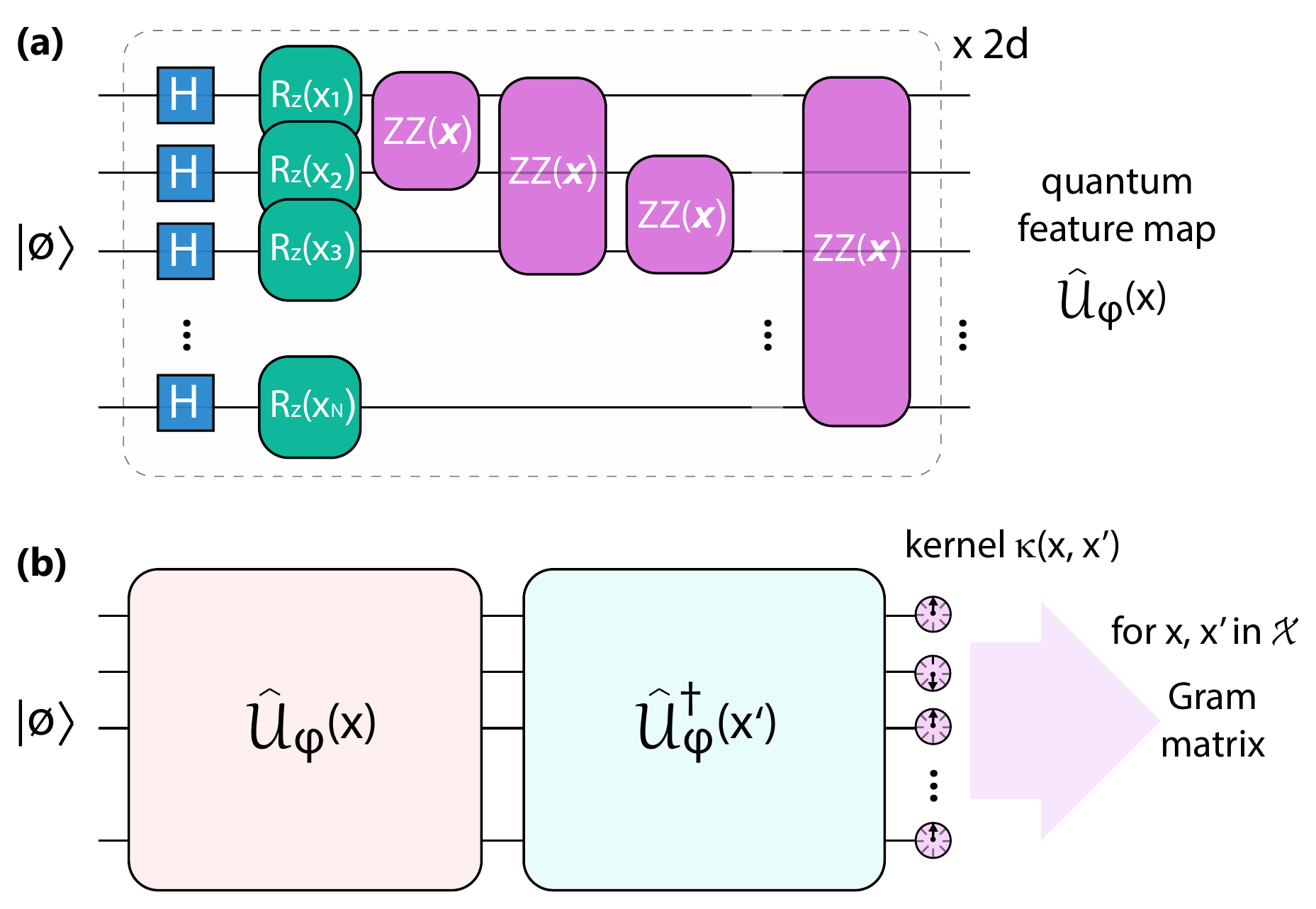}
\caption{\textbf{Quantum kernel circuits. (a)} Quantum feature map $\hat{\mathcal{U}}_\phi(\mathbf{x})$ for embedding data sample $x \in \mathcal{X}$ based on IQP-like circuit. It consists of a layer of Hadamard gates, followed by feature-dependent Z-basis rotations and ZZ gates between all qubits. Single IQP embedding corresponds to a twice-repeated circuit, and we consider up to $d=3$ re-uploadings. \textbf{(b)} Simple qubit-frugal measurement scheme for quantum kernels as a probability of measuring initial state $|{\o}\rangle$ after the action of $\hat{\mathcal{U}}_\phi(\mathbf{x})$ and reversed circuit $\hat{\mathcal{U}}_\phi(\mathbf{x}')$ for a modified sample. Repeating measurement for various points in the dataset allows assembling the Gram matrix.}
\label{fig:kernels}
\end{figure}
%%%

\textbf{Unsupervised anomaly detection.} Once the supervised benchmarks are established, we build a classical workflow for unsupervised anomaly detection. Here, we concentrate on kernel-based methods and specifically the one-class support vector machine approach. The idea is to separate a class of nominal points by a hyperplane such that they are distinct from anomalous samples. This is also sometimes referred as novelty detection \cite{Scholkopf1999}. Similarly to SVC, OC-SVM relies on calculating Gram matrices for datasets with confirmed (nominal) transactions. Additionally to kernel-related hyperparameters, OC-SVM relies on another hyperparameter $\nu$ being an estimated fraction of anomalies. We stress that the knowledge about classes of transactions $\mathbf{Y}$ is only used when plotting the average precision, but not used when building models.

The results for OC-SVM anomaly detection with RBF kernels are shown in Fig.~\ref{fig:classical}(b), as a function of increased feature space (number of principal components). Here, we used the default values suggested for OC-SVM taken as $\gamma = 1 / (N \mathrm{Var}[\mathcal{X}])$ and small maximum anomaly fraction $\nu = 0.1$. We observe a gradual increase of average precision for anomaly detection at growing number of features, that starts to saturate around 15 features at an AP of 0.55. Hyperparameter tuning of $C$ and $\gamma$ does not yield significantly different results. A small decrease of AP for large $N$ can be seen as an inability to use correlations from remaining principal components. As quantum kernels offer expressivity advantage and believed to be suitable for highly-correlate datasets, we test the unsupervised OC-SVM below using quantum data encoding.

%%% ========= QUANTUM =========== 

\section{Fraud detection: quantum approach} We proceed to develop the protocols for quantum anomaly detection. Given the connection between classical and quantum kernels \cite{Schuld2021kernel}, we propose to use unsupervised learning approaches with the kernel structure, and employ quantum kernels. We start with a dataset of points $\{ \bm{x} \} \in \mathcal{X}$ in the input space $\mathcal{X}$. First, we need to embed the data into a quantum state $|\phi(x)\rangle$ of an $N$-qubit register. We approach this by performing a feature map as a quantum circuit $\hat{\mathcal{U}}_\phi(x)$ applied to the trivial product state, $|\phi(x) \rangle = \hat{\mathcal{U}}_{\phi}(x) | {\o} \rangle$. One example, corresponding to the instantaneous quantum polynomial (IQP) circuit is shown in Fig.~\ref{fig:kernels}(a). This maps a datum $x$ into a high dimensional space, facilitating the use of kernel trick as employed in support vector machines, where the (generally nonlinear) map $\phi$ modifies and often increases the space dimension of data (leading to so-called \emph{latent space} representation). The quantum feature map is defined by the circuit structure $\hat{\mathcal{U}}_\phi(x)$, where the function $\phi$ determines gate properties. Additionally, we may use a variational circuit $\hat{V}(\bm{\theta})$ (defined by its structure and a vector of variational parameters $\bm{\theta}$). This adapts the measurement basis for the readout, and can be used later for the quantum kernel alignment \cite{Hubregtsen2021,Glick2021}. Otherwise, fixed circuit parameters are used. The sequence of circuits $\hat{V}(\bm{\theta}) \hat{\mathcal{U}}_\phi (x)$ is often called a quantum neural network (QNN). It generates a high-dimensional latent space representation of the data point $x$. 
The choice of a feature map defines the success of learning procedure, determined by its \emph{expressivity} \cite{Schuld2019QML,SchuldSweke2021,Abbas2021,kyriienko2021generalized}.

For running a quantum protocol we use a quantum kernel trick for support vector machines (SVM). Quantum SVM operation relies on the comparison of data points in the Hilbert space. This can be measured as an inner product $\kappa(x, x') = |\langle \psi(x) | \psi(x') \rangle|^2$. The overlap can be measured in a simple form by evaluating return probabilities [Fig.~\ref{fig:kernels}(b)], albeit requiring an increased number of measurement shots. Other options for overlap-based kernel evaluation include SWAP and Hadamard tests \cite{schuld2019evaluating}, that use ancillary qubits and controlled gates. Alternatively, one may substitute state overlap by an effective distance measurement $\kappa_{\hat{P}}(x, x')$, where expectations of local Pauli operators $\{ \hat{P}_k \}_{k=1}^N$ are compared instead. This is referred as the projective kernel method \cite{Huang2021}, which led to classification with up to thirty qubits when using high-performance computing. Recently, it was shown that under conditions of controlled kernel bandwidth overlap-based and projected quantum kernels lead to similar performance \cite{Shaydulin2021}, and it is possible to choose one that is more suitable to particular system. Finally, once the Gram matrix is evaluated on a quantum computer (or its simulator), the model is formed using the same steps as in the classical workflow.
%%%
\begin{figure}
\centering
\includegraphics[width=1.\linewidth]{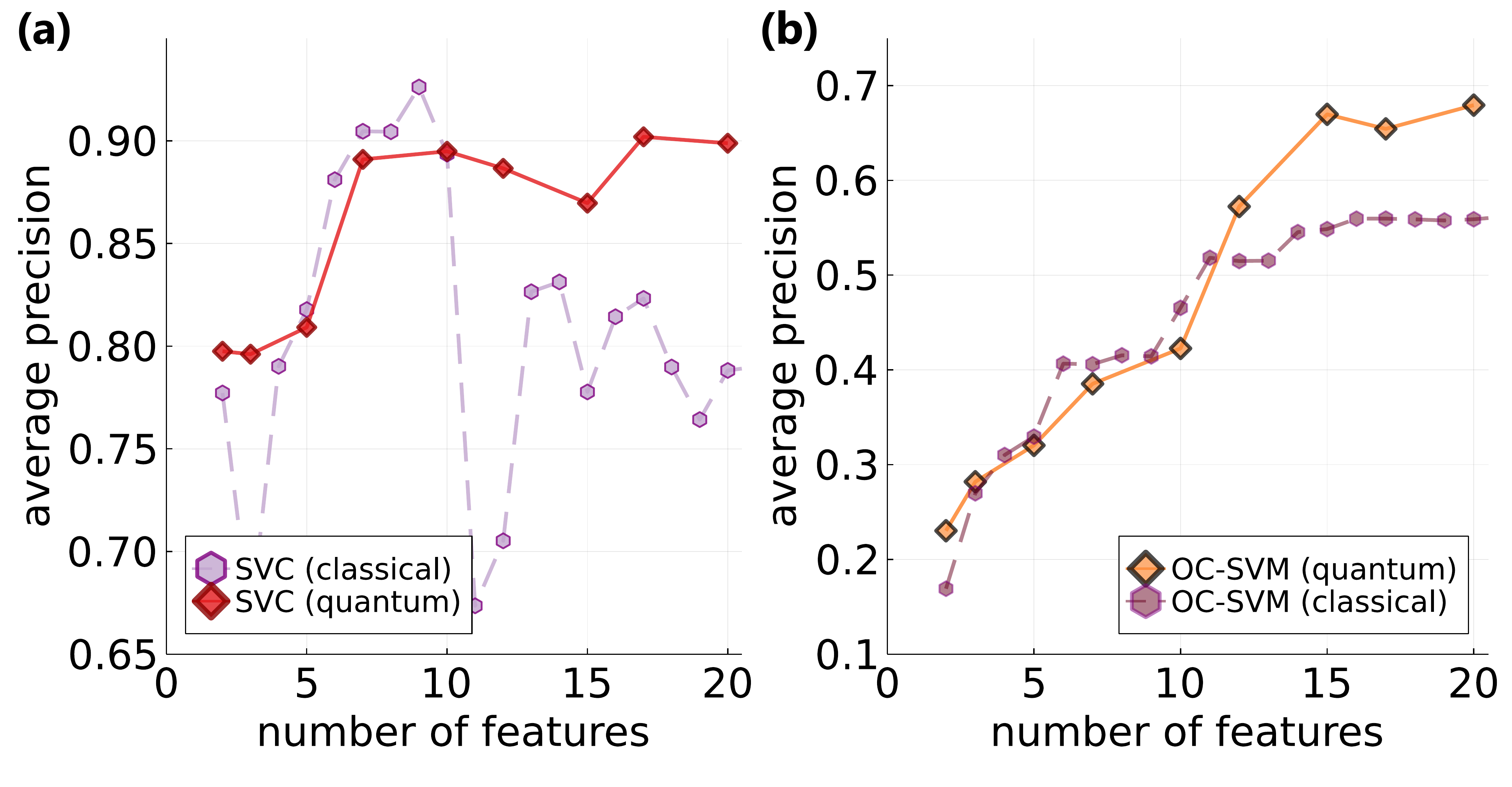}
\caption{\textbf{Quantum fraud detection. (a)}. Average precision (AP) of supervised QML based on quantum kernels, shown at increased number of qubits (features). AP shows the separation between quantum and classical performance at more than 10 qubits. \textbf{(b)} Average precision of unsupervised QML based on OC-SVM with IQC quantum kernels, shown at increased number of qubits (features). We observe clear quantum-classical separation, where IQP kernel with $d=3$ re-uploading reaches as large as $15\%$ higher AP at 20 qubits.}
\label{fig:quantum}
\end{figure}
%%%

We proceed to choose possible quantum feature maps. Concentrating on the near-term operation, these typically correspond to either products of data-dependent rotations with variational circuits in between \cite{Mitarai2018,kyriienko2021solving}, evolution circuits with Heisenberg or spin-glass Hamiltonian \cite{Huang2021,Henry2021}, or IQP circuits that lead to computationally-hard kernels \cite{Havlicek2019,Coyle2020}. Other options with proven exponential advantage come from circuits used for solving discrete logarithm problems \cite{Liu2021}. Each feature map can be applied several times following the same circuit structure (so-called data re-uploading technique \cite{PerezSalinas2020datareuploading,SchuldSweke2021}), increasing the expressivity of circuits. Unless some nonlinear map is applied to $x$, the associated quantum kernels can be seen as Fourier series $\kappa(x,x') = \sum_{\omega \in \Omega} c_{\omega} \exp[-i\omega (x - x')]$, where a set frequencies $\Omega = \{\omega\}$ is defined by generators of data encoding, and may contain up to exponentially many frequencies \cite{Kyriienko2022}. The coefficients of Fourier series $c_\omega$ are defined by data-independent circuits. We note that given the periodic nature of quantum encoding, it is important to normalize features such that data samples $x \in \mathcal{X}$ are distinguished without ambiguity.   

We perform numerical test for quantum kernels at small system sizes, and observe that overall IQP circuits with re-uploading offer highest performance for classification. We keep the number of qubits being equal to the number of features, where one qubit is responsible for uploading one feature, and the information is spread by quantum circuits that scramble the state. Specifically, we concentrate on IQP circuits re-uploaded $d=3$ times, such that mappings are expressive, yet do not lead to overfitted models. The structure of circuits is described in Refs. \cite{Havlicek2019,Huang2021,Shaydulin2021}, and is shown in Fig.~\ref{fig:kernels}(a). IQP mapping consists of a layer of Hadamards, followed of $x$-dependent rotations around $X$ axis of each qubit (one feature per qubit). Then, we evolve states under pairwise two-qubit interactions of feature type, with phases being dependent on products of two features. The sequence is repeated twice to form the IQP layer (per convention), and re-uploaded $d=3$ times with shallow random quantum circuits in between to avoid coherent gate cancellation. We do not include any trainable parameters in the quantum circuit. The kernels are evaluated using return probabilities, and we use state vector simulation.

We proceed to run quantum kernel classification and anomaly detection. The same credit card dataset is used, with each feature being scaled by a prefactor $\eta = 0.1$. This allows to make sure that phases of embedding are all smaller than $2\pi$, and naturally lowers the bandwidth of quantum kernels. We also confirmed that fine-tuning is not necessary and modification of $\eta$ does not lead major performance changes. For defining and simulating the quantum circuits, we used \textsf{Pennylane} library for Python, with the JAX interface that allows for compiling the circuits with XLA. This results in fast, parallelized operation across CPU and GPU resources. Additionally, we implement batched Gram matrix evaluation with kernel evaluations streamlined over different values. For computing we used an NVIDIA A100 GPU with 40~GB memory (peak memory bandwidth up to 1555 GB/s). 

The results of supervised quantum classification are shown in Fig.~\ref{fig:quantum}(a), where IQP kernel performance is shown in red diamonds (solid curve). At small number of features we observe increase of average precision up to 0.9 values for five-to-ten features. Notably, the classical kernel model performance deteriorates as we go beyond 10 features, while the quantum kernel model performance does not, and even reaches new highs around 17 features/qubits. At $N=20$ we see a clear separation of performance between quantum and classical (RBF) kernels. This suggests that quantum kernels avoid overfitting, a feature previously studied in other works \cite{Caro2021,Caro2021arxiv}. It can be also interpreted as quantum mappings being less sensitive to noise added with the higher order principal components. This is a positive sign showing expressivity/learning advantage for certain tasks performed on classical datasets. 

Next, we proceed to describe findings from unsupervised quantum anomaly detection. The results are shown in Fig.~\ref{fig:quantum}(b). Similarly to classical RBF kernels, quantum kernels based on IQP circuits show improved average precision, with a separation developing past $N=10$ features (qubits). Importantly, AP for quantum kernels continue to grow, and show maximal AP of up to 0.7 at $N=20$ qubits. This shows that quantum kernels outperform classical kernels with additional 15\% of precision, showing a favourable operation of quantum kernels. This suggests a potential use of quantum kernels when working with datasets that require learning for correlated multi-feature data in the unsupervised fashion. This represents one of the main messages of the study.

%%% ========= DISCUSSION ===============

\section{Discussion and perspectives} We see that quantum kernels can outperform classical kernels at increasing size of feature space, and show their excellent suitability for unsupervised anomaly detection. Does it make QC a viable choice for industrially-relevant fraud detection? Let us critically analyse potential benefits and drawbacks for future applications of quantum computers in the sector. 

As an obvious benefit we consider an increased quality of detecting fraud --- even a few-percent increase of average precision can lead to multi-million savings (with a billion pounds of fraud prevented yearly in the UK, changes in detection mechanism of 15\% bring £150 million back into economy). Another important feature of developing unsupervised methods is in detecting potentially new types of fraud (unlike simple classification), working towards preventing cases that are not tracked with current methods. To drive the change of detection mechanisms and adopt QC-based (or generally ML-driven) strategies we need to assess the required resources. Specifically, to understand future steps in QC for fraud detection we need to: 1) define relevant timescales and precision targets for training and inference of fraudulent transactions; 2) match these timescales with resources available currently and foreseen in the near- and mid-term future.

Training of FD models based on quantum kernels crucially depends on the ability to evaluate the Gram matrix of distances between $N_s$ samples (e.g. transactions). Typical sizes of credit card datasets count over 100,000 representative samples (further subsampled for scalable training). Given the $O(N_s^2)$ scaling of the full Gram matrix $\mathbb{K}$ estimation in terms of kernel evaluations, the computational cost can be large. This shall be ideally parallelized (using different QC nodes) and in principle can take significantly long time for building models. In the case of constantly appearing new transactions $\mathbb{K}$ needs to be updated on-the-fly. The corresponding cost is $O(N_s)$ in terms of kernel evaluations for adding a sample, which is substantial when working with large datasets. Important training time scales in this case can be at the scale of weeks (or months) during which the fraud landscape evolves, giving enough time for adjusting the fraud detection models. We note that this part is similar to large-scale training of deep neural networks, that despite large computational cost for training (several weeks and more on multi-GPU clusters), offers an excellent way for improving performance. We expect that relevant tasks in fraud detection with quantum computers will require similar training times, matching the pace at which new types of frauds emerge. However, a crucial point is not just in building models, but the time needed for inference. Namely, to assign a label (fraud or not) for detecting $N_d$ new-coming samples we require $O(N_d N_s)$ kernel evaluations. This requires significant resources if all incoming transactions are tested, and can be hindered by relatively slow absolute processing rates of quantum hardware (see the discussion below). If a single entry can be tested with sub-millisecond operational time one can consider it for real-time use. However, if the test requires an hour, this automatically precludes QC runs for real-time detection, and can only be used during additional checks. Thus, QC-based fraud detection may be not well-suited for \emph{en masse} testing of transactions, but rather checks of highly prioritised cases. We believe that the question of prioritisation and reduction of kernel evaluations (choosing those points that lead to non-negligible overlaps) is one of fundamental questions of quantum fraud detection.

Now, let us match the required timescales to currently available quantum resources (we set them as weeks for training and as short as possible for inference). To do the analysis, we choose three different platforms: 1) superconducting circuits (SCs); 2) trapped ions; 3) optical systems. Importantly, in each case we do not stick to concrete realisations, but rather analyse possible best and median performance as derived from physical limits of each system. For the reduced size dataset (as considered in numerical statevector calculations) we use 500 samples, with the size of Gram matrix being $\sim 100,000$. The quantum cost of kernel evaluation is then defined as circuit runtime $\tau$ times the number of measurement shots $N_{\mathrm{shots}}$ needed to collect the statistics. We also consider training and inference times for large datasets that arise in industrially-relevant use-cases, setting the number of samples to $\sim 100,000$.

\textbf{Superconducting circuits.} First, we consider SC chips, which are developed by various companies and academic groups worldwide  \cite{Kandala2017,Jurcevic2021,Arute2019,Satzinger2021,Wu2021,Patterson2019,Krinner2022}. This powerful quantum platform has shown tremendous growth due to the mixture of good scalability and microwave-frequency operation. With typical quantum gate times ranging in 10-100~ns \cite{Jurcevic2021,Arute2019,Satzinger2021}, and T$_1$ times of $\sim 10~\mu$s, single circuit evaluation may be performed within a microsecond, leading to MHz repetition. More conservative estimates correspond to 1~kHZ repetition, where a passive reset is used \cite{Corcoles2021}. Reproducible kernel measurements require at least $N_{\mathrm{shots}} = 10^3$ measurement shots, and for high-quality estimates at increased register size this may grow to $10^6$ (e.g. typical budget for near-term quantum chemistry protocols \cite{Kubler2020}). As we require $10^5$ kernel evaluations, this brings us to $100$~s training time at optimistic MHz rates for reduced-size datasets and minimal number of shots. If number of shots grows to $10^6$, the training time grows to 28 hours with a single QC. For the pessimistic KHz gate rates, the minimal budget of $10^3$ matches the same cost of 28 hours. The inference time is significantly smaller in the case of reduced dataset, as it takes $0.5$~s for minimal shots (can enable close to real time detection), and increases into minuted for $10^6$ shots. This poses an important question of $N_{\mathrm{shots}}$ required for stable estimation of Gram matrices and high-quality inference.

For large datesets of $\sim 100,000$ samples forming full Gram matrices is an extremely costly procedure. With MHz absolute rates and $10^3$ shots the full calculation can require 16 weeks. However, one may potentially use randomized protocols for a \emph{partial} inference of Gram matrices, and adaptive detection \cite{Haug2021}. In this case the asymptotic scaling of training becomes linear with the number of samples, $O(N_s)$, reducing its time to $\sim 1$~min, with potentially much smaller inference times. We note that the task of fraud detection with small training and inference times is a priority for future work and implementation.

\textbf{Trapped ions.} Next, we consider trapped ions platform with high fidelity gates but slower absolute clock rates of quantum processors. With $1~\mu$s execution time for a single gate \cite{Egan2021,Hughes2020}, as well as long T$_1$ (up to second), the runtime $\tau$ for a sequence of gates optimistically reaches $100~\mu$s (including costly readout), and absolute rates per shot of 10~kHz. Taking the 500-sample dataset with $10^5$ kernel evaluations and assuming $N_{\mathrm{shots}} = 10^3$ measurement shots, we require around 3 hours for training. For $N_{\mathrm{shots}} = 10^6$ full training requires 17 weeks, calling for shot-frugal kernel evaluation and parallel kernel evaluation.

For large datasets of $N_s = 100,000$ samples forming full Gram matrices at 10 kHz rates may become infeasible with a single processor, as a brute force evaluation even at $N_{\mathrm{shots}} = 10^3$ may require 30 years, with 3 hours of inference time. This suggests that a drastic reduction of the number of kernel evaluations (lowering down scaling to linear) and number of shots is required for working with big data.

\textbf{Photonic chips.} Finally, we consider potential optical realizations of quantum computers. Hypothetically, these shall be characterized by highest absolute clock rates of quantum processors, defined by photo lifetime in photonic structures reaching 100~ps \cite{Nelsen2013,Liu2016,Kuriakose2022}. With gate sequence times of comparable size, performing $N_{\mathrm{shots}} = 10^3$ makes training of small dataset fast (10 milliseconds) that grows into 10 seconds for training with $N_{\mathrm{shots}} = 10^6$. The large $100,000$-sample dataset can take 17 minutes of training at $N_{\mathrm{shots}} = 10^3$, remaining manageable for more shots if the sparse Gram matrix is used. However, we note that there is no established way to do optical computing with deterministic gates yet, so finding suitable architectures shall be prioritized when measuring large number of shots is required.

%\textbf{Recommendations and future directions.} We collect our analysis into proposal for several directions crucial for future QFD implementation:
%\begin{itemize}
    %\item Quantum kernels used for fraud detection shall be tuned in bandwidth (dataset re-scaling) to over high performance, and ideally have structure representing correlations in each dataset. Here, possible problem becomes data leakage --- once we train model in supervised way, we may hinder unsupervised anomaly detection via additional bias. 
    %\item The size of datasets is typically large, and quadratic scaling for training requires tailored ways of forming a Gram matrix (randomized or sparse approaches rather then bruteforce full evaluation). This task is even more crucial for quantum hardware with smaller absolute rates of operation.
    %\item Number of measurement shots needs to be established at the level where kernel measurement is close to deterministic (possibly, controlled by kernel structure), yet $N_{\mathrm{shots}}$ is minimized to reduce runtimes.
    %\item One of the main challenges we see the increase of inference time for large datasets/many shots/slow absolute clock times. 
%\end{itemize}

%%% ============= CONCLUSIONS ===================

\section{Conclusions} We studied the problem of fraud detection based on unsupervised quantum machine learning. For this we developed protocols that utilize kernel structure (specifically, one-class support vector machines) and tested anomaly detection with different types of quantum kernels. Using the credit card fraud dataset with PCA pre-processing, we tested the approach at increasing number of features (qubits), and compared to fine-tuned classical methods. Our numerical simulations ranged from few to 20 qubits, where we used fast batched simulation run on NVIDIA A100 GPU. For the supervised fraud classification task we observed that quantum kernels offer higher expressivity and generalization, outperforming RBF kernels by over ten percent of average precision (AP). Moreover, for unsupervised fraud detection we observed a clear separation between average precision of quantum kernel methods and classical unsupervised learning, showing 15\% increase of AP that grows at larger system size. Given favourable scaling \textit{in silico}, we analysed possible timescales for future quantum hardware simulations.

We note several directions for future research that may appear crucial for further developing the quantum fraud detection. First, quantum kernels used for fraud detection can be adjusted to optimize the performance, ideally having a structure that represents correlations in each dataset. Here, a possible problem becomes data leakage --- once we train model in supervised way, this may hinder unsupervised anomaly detection via additional bias. Second, as size of datasets is typically large, the quadratic scaling for training times shall be lowered. This requires tailored ways of forming a Gram matrix, possibly with randomized or sparse approaches rather than the bruteforce full evaluation. This task is even more crucial for running on quantum hardware with smaller absolute rates of operation. Third, the number of measurement shots needs to be established at the level where kernel measurement is close to deterministic (possibly, controlled by kernel structure), yet $N_{\mathrm{shots}}$ is minimized to reduce the runtime. %As one of the main challenges for quantum FD we see the increase of inference time for large datasets/many shots/slow absolute clock times.

Let us also comment on the potential for reaching a quantum advantage for considered use-case and approaches. Additionally to the discussion of quantum hardware capabilities, we stress that the timeline for QC-based fraud detection largely depends on insimulability of the proposed approach on classical hardware. Namely, advantageous implementation of quantum FD requires going beyond calculations performed for 20-qubit registers (reaching 50-100 qubits for strong separation), as well as deep quantum kernels. We also note the requirement for fast sampling and low latency operation, which can potentially offer quantum advantage. The exact timeline of implementing quantum FD in industrial pipelines largely depends on the combination of discussed factors and ability to solve open problems.

\subsection*{Disclaimer} This paper was prepared for information purposes, and is not a product of HSBC Bank Plc. or its affiliates. Neither HSBC Bank Plc. nor any of its affiliates make any explicit or implied representation or warranty and none of them accept any liability in connection with this paper, including, but limited to, the completeness, accuracy, reliability of information contained herein and the potential legal, compliance, tax or accounting effects thereof. This document is not intended as investment research or investment advice, or a recommendation, offer or solicitation for the purchase or sale of any security, financial instrument, financial product or service, or to be used in any way for evaluating the merits of participating in any transaction.\\

\begin{acknowledgments}
We would like to thank Omar Jamil and Felicity Guest for helping with classical benchmarking. The authors acknowledge the support from Innovate UK ISCF Germinator project number 10003408.
\end{acknowledgments}

%%%%%%%%%%%%%%%%%%%%%%%%%%%%%%%%%%%%%%%
%Bibliography
%%%%%%%%%%%%%%%%%%%%%%%%%%%%%%%%%%%%%%%

\vspace{10pt}

%\printbibliography

\bibliography{Bibliography}

\end{document}